\documentclass[twoside,reprint,superscriptaddress,amsmath,amssymb,aps]{revtex4-1}
\usepackage[latin9]{inputenc}
\usepackage{color}
\usepackage{float}
\usepackage{amsmath}
\usepackage{amssymb}
\usepackage{graphicx}
\makeatletter
 
 \@ifundefined{textcolor}{}
 {%
   \definecolor{BLACK}{gray}{0}
   \definecolor{WHITE}{gray}{1}
   \definecolor{RED}{rgb}{1,0,0}
   \definecolor{GREEN}{rgb}{0,1,0}
   \definecolor{BLUE}{rgb}{0,0,1}
   \definecolor{CYAN}{cmyk}{1,0,0,0}
   \definecolor{MAGENTA}{cmyk}{0,1,0,0}
   \definecolor{YELLOW}{cmyk}{0,0,1,0}
 }

\usepackage{dcolumn}
\usepackage{bm}

\def\be{\begin{eqnarray}}
\def\ee{\end{eqnarray}}

\makeatother

\begin{document}

\title{Effect of long-range spatial correlations on the lifetime statistics of an emitter in a two-dimensional disordered lattice}

\author{N. de Sousa}

\affiliation{Departamento de F\'isica de la Materia Condensada, Universidad Aut\'onoma
de Madrid, 28049, Madrid, Spain.}

\author{J. J. S\'aenz}

\affiliation{Departamento de F\'isica de la Materia Condensada, Universidad Aut\'onoma de Madrid, 28049, Madrid, Spain.}

\affiliation{Condensed Matter Physics Center (IFIMAC) and Instituto ``Nicol\'as Cabrera'', Universidad Aut\'onoma
de Madrid, 28049, Madrid, Spain.}

\affiliation{Donostia International Physics Center (DIPC), Paseo Manuel de Lardizabal 5, Donostia-San Sebasti�n 20018, Spain.}

\author{A. Garc\'ia-Mart\'in}

\affiliation{IMM-Instituto de Microelectr\'onica de Madrid (CNM-CSIC), Isaac Newton
8, PTM, Tres Cantos, E-28760 Madrid, Spain.}

\author{L. S. Froufe-P\'erez}

\affiliation{Instituto de Estructura de la Materia, (IEM-CSIC), Serrano 121, 28006
Madrid, Spain.}

\author{M. I. Marqu\'es}
\email[]{manuel.marques@uam.es}

\affiliation{Condensed Matter Physics Center (IFIMAC) and Instituto ``Nicol\'as Cabrera'', Universidad Aut\'onoma
de Madrid, 28049, Madrid, Spain.}

\affiliation{Departamento de F\'isica de Materiales, Universidad Aut\'onoma
de Madrid, 28049, Madrid, Spain.}

\begin{abstract}
The effect of spatial correlations on the Purcell  effect in a bidimensional  dispersion of resonant nanoparticles is analyzed. We perform extensive calculations on the fluorescence decay rate of a point emitter embedded in a system of nanoparticles  statistically distributed according to a simple  2D lattice-gas model near the critical point.  For short-range correlations (high temperature thermalization) the Purcell factors present a long-tailed statistic which evolves towards a bimodal distribution when approaching the critical point where  the  spatial correlation length diverges.   Our results suggest long-range correlations as a possible origin of the large fluctuations of experimental decay rates  in disordered metal films.
\end{abstract}


\maketitle
Since Purcell's work \cite{Purcell_1946}, it is known that the lifetime
of an excited atomic state is a combination of the atom's properties
and the environment where it is embedded. Changes in the emission
decay rate in regular structures have been reported for emitters placed
close to planar interfaces \ \cite{Ford_1984,Chance_1978,Waldeck_1985,Barnes_1998},
cavities \cite{Berman_1994}, photonic crystals \cite{Lodhal_2004}, plasmonic \cite{Carminati_OpCom_2006, Castanie_2012_distance} and magnetoplasmonic structures \cite{Nikolova_2013},  
among others. More recently, the possibility of creating nanostructured materials stimulated the interest in waves
propagation through disordered media \cite{Wiersma_2013}. Examples like backscattering enhancement \cite{Ishimaru_1985,Mark_1998},
photon localization \cite{Wiersma_1997} random lasing \cite{Lawandy_1994,Cao_2000}, or photonic membranes \cite{Caze_2013,Garcia_2012} can be found in the literature. Also, the behavior of light coupled to disordered matter has been analyzed from the point of view of diluted cold atom systems in scarcely populated optical lattices \cite{Froufe_2009_rlaser, Bloch_2012}.
In complex systems like liquids, colloids, granular or biological
materials, the dynamic modification of the environment or the movement
of the emitter implies the need for a statistical study of the decay rate \cite{Froufe_PRA_2007,Froufe_2008}. In these random systems, the decay rate exhibits a non Gaussian, long-tailed distribution,
where large enhancements of the Purcell factor are attributed to strong fluctuations in the local density of states induced by the near-field scattering
\cite{Martorell_1991,Krachmalnicoff_2010,Sapienza_2011}.
These rare events create optical modes confined in a small volume around the source. 

In most cases, previously-studied disorder was produced in a random
way, that is, scatterers were distributed throughout the lattice randomly.
Real systems, however, can be realized with other kinds of disorder,
where the locations of the dipoles are correlated. In particular,
structural disorder with long-range correlation (LRC) has been found
for instance in X-ray and Neutron Critical Scattering experiments
in systems undergoing magnetic and structural phase transition \cite{ShiraneI,ShiraneII,ShiraneIII,ShiraneIV}.
This correlation effect in magnetic systems may be modeled by assuming
a spatial distribution of critical temperatures with a correlation function obeying a slow power law
decay \cite{Weinrib,Altarelli,BallesterosIV}.
Spatial correlations in disordered scattering materials have been shown to dramatically modify the light transport properties, in particular, light scattering mean free path presents strong chromatic dispersion \cite{Rojas_2004, Froufe_2009}.

In the present paper, long-range correlated distributions of scatterers
are produced using a thermal order-disorder distribution governed
by a characteristic ordering temperature ($\theta$) \cite{Marques,MarquesII}.
In particular, the quenched randomness is implemented using a lattice-gas model equivalent to a thermally-diluted
ferromagnetic two dimensional (2D) Ising lattice at a temperature
$(\theta)$. We perform extensive calculations of the fluorescence decay rate of a point emitter embedded in a system of nanoparticles  statistically distributed according to a 2D lattice-gas model near the critical point.  As we will show, for short-range correlations (high temperature thermalization), the Purcell factors present a non-Gaussian long-tailed statistic where events with large Purcell factors are extremely rare. Interestingly, as we approach the critical point where  the  spatial correlation range diverges, the statistics evolves towards a bimodal distribution with a well-defined peak at high enhancement factors. Our numerical results strongly resemble those obtained experimentally in resonant thin metallic films near percolation \cite{Krachmalnicoff_2010},
 which suggest  long-range correlations as a possible origin of the large fluctuations of experimental decay rates  in disordered metal films. 

\begin{figure}[H]
\begin{centering}
\includegraphics[width=8cm]{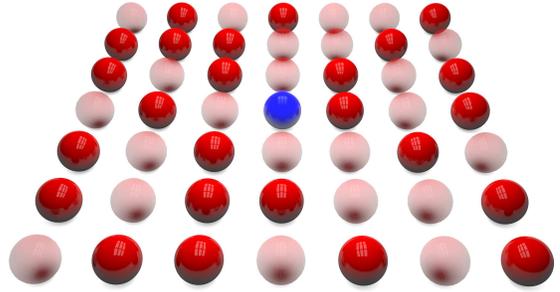} 
\par\end{centering}
\caption{(Color online) Schematic representation of a crystalline structure with an edge ($D$)
of 7 particles. The transparent spheres represent the removed particles. The dipole emitter is represented by the blue sphere
and the scatterers by red spheres.\label{fig:crystalline_structure}}
\end{figure}

Let us consider the lattice system sketched in Fig. (\ref{fig:crystalline_structure}).
The nodes of this lattice are taken as the possible location
points of the optical dipoles by taking into account the following
mechanism: After thermalization of the pure Ising model at temperature
$\theta$, the spins with $s=1$ are taken as the locations of the
scattering dipoles, while spins with $s=-1$ are considered as dipole
vacancies (sites with no optical response). The structure of the realization
so constructed is fixed thereafter for all subsequent optical decay
rate studies.

In these systems, the correlation function between point dipoles separated by a distance $r$  is  given by the spin-spin correlation function \cite{Yeomans}:

\begin{equation}
g(r)\sim r^{-\tau}exp^{-r/\xi}
\end{equation}

$\xi$ being the correlation length and $\tau$ a characteristic number depending on the system. The correlation length, near the critical temperature 
of the Ising model ($T_ {c}$), depends on the ordering temperature with a power law given by $\xi\sim(\theta-T_{c})^{-\nu}$, where $\nu=1$ and  
$T_{c}=2/log(1+\sqrt{2})$ for the two-dimensional Ising system.

So, if a high ordering temperature ($\theta>>T_{c}$) is used to generate a particular realization,
the correlation function of the equilibrium thermal disposition of the dipoles
is going to be given by $g(r)\sim exp(-r/\xi)$, i.e., we have a random (short-range correlated)
disorder similar to the one used in previous investigations \cite{Froufe_2008, Martorell_1991,Sapienza_2011}. However, if $\theta$ happens
to coincide with the characteristic critical temperature
of the pure Ising model ($\theta=T_{c}$) then $\xi \rightarrow \infty$ and we are going to have
scatterers randomly located, but following a long-range correlated
distribution given by $g(r) \sim r^{-\tau}$. 
The dispersion on dipole concentration ($c$) obtained by a thermal
distribution at $\theta=T_{c}$ is larger than the one obtained when scatterers are randomly
distributed with no correlation. In order to be able to compare both
cases we only consider systems with a concentration in the range between
$c=0.45$ and $c=0.55$. 
More details about the construction
of these thermally-disordered systems can be found in refs. \cite{Marques,MarquesII}.

We consider a square lattice with lateral size $D$, where the emitter
is placed in the central position of the lattice and oriented out
of plane (see Fig. (\ref{fig:crystalline_structure})). 

In this paper we consider a particular frequency ($\omega = \omega_{0}$) and an associated particular wavenumber ($k = k_{0}=\omega_{0}/c$) at which dipoles are in resonance with the electromagnetic radiation, meaning that the polarizability is now given by $\alpha = i6\pi/k_{0}^3$. 
In the presence of a dipole emitter $\mathbf{p}(\mathbf{r})$ the electric field at some position $\mathbf{r'}$ can be obtained by operating the Green tensor over the
dipole positioned at $\mathbf{r}$. Mathematically, this is expressed
as: 
\begin{equation}
\mathbf{E}\left(\mathbf{r}'\right)=\frac{k^{2}}{\epsilon_{0}}\mathbf{G}_{0}\left(\mathbf{r}',\mathbf{r}\right)\cdot\mathbf{p}(\mathbf{r}),\end{equation}
$\epsilon_{0}$ being the permittivity of vacuum.

The Green tensor is given by \cite{Novotny_PNO_2012}: 
\begin{equation}
\begin{split}\mathbf{G}_{0}\left(\mathbf{r},\mathbf{r}'\right)= & \frac{e^{ikR}}{4\pi R}\left[\left(1+\frac{ikR-1}{k^{2}R^{2}}\right)\mathbb{I}+\right.\\
 & \left.+\left(\frac{3-3ikR-k^{2}R^{2}}{k^{2}R^{2}}\right)\hat{\mathbf{R}}\otimes\hat{\mathbf{R}}\right],
\end{split}
\end{equation}
where $R$ is the modulus of the vector $\mathbf{R}=\mathbf{r}-\mathbf{r}'$, $\hat{\mathbf{R}}\otimes\hat{\mathbf{R}}$
denotes the outer product of $\hat{\mathbf{R}}=\mathbf{R}/R$ by itself and $\mathbb{I}$ is the unit dyadic.

When the emitter is in the presence of $N$ dipole scatterers,
the scattered field at position $\mathbf{r'}$ is given by: 
\begin{equation}
{\mathbf{E}\left(\mathbf{r'}\right)=\frac{k^{2}}{\epsilon_{0}}\mathbf{G}_{0}\left(\mathbf{r},\mathbf{r}'\right)\mathbf{p}(\mathbf{r})+\atop +\frac{k^{2}}{\epsilon_{0}}\sum_{{m=1}}^{N}{\mathbf{G}_{0}\left(\mathbf{r},\mathbf{r}_{m}\right)}\mathbf{p}_{m},}\label{eq:electric_field}
\end{equation}
where $\mathbf{r}_{m}$ is the position of the $m$ scatterer and $\mathbf{p}_{m}=\epsilon_{0}\alpha\mathbf{E}(\mathbf{r_{m}})$ is the value of the induced dipole located at $\mathbf{r_{m}}$. To obtain the value of all induced dipoles we should solve the electric fields in Eq. (\ref{eq:electric_field}) by considering the coupled dipole method \cite{Purcell_1973}.
The second term on the right-hand side of Eq. (\ref{eq:electric_field})
represents the modification of the free-space dyadic Green function
due to the presence of the scatterers. The scattered field is then given by:
\begin{equation}
\mathbf{E}_{s}\left(\mathbf{r}'\right)=k^{2}\alpha\sum_{{m=1}}^{N}{\mathbf{G}_{0}\left(\mathbf{r}',\mathbf{r}_{m}\right)}\mathbf{E}(\mathbf{r_{m}}).\label{eq:scattering_field}
\end{equation}

Once the scattered field is known, the normalized spontaneous decay rate
$\Gamma$ of a dipole $\mathbf{p}(\mathbf{r'})$, in the weak coupling regime, is given by \cite{Novotny_PNO_2012}:
\begin{equation}
\frac{\Gamma}{\Gamma_{0}}=1+\frac{6\pi\epsilon_{0}}{\left|\mathbf{p}\right|^{2}k^{3}}\Im[\mathbf{p}^*\cdot\mathbf{E}_{s}\left(\mathbf{r}\right)],\label{decay_rate}
\end{equation}
where $\Im$ is the imaginary part and $\Gamma_{0}$ is the decay rate of the emitter in free space. It is important to take into account that the expression for the spontaneous decay rate we are considering is an approximation valid only for weak interacting fields. In fact, Eq. (\ref{decay_rate}) is obtained by quantum electrodynamics calculations of the spontaneous decay rate of an atomic system in an inhomogenous medium when the weak coupling regime approximation is considered \cite{Novotny_PNO_2012}.

First, we analyse the full crystalline configuration and calculate the
normalized decay rate of the emitter, positioned in the center of
the structure, as a function of the ratio between the lattice parameter
$a$ and the resonance wavelength considered $\lambda=\frac{2\pi}{k}$. We fix the wavelength of the emitter and we vary the particles' position in the lattice.

\begin{figure}[H]
\begin{centering}
\includegraphics[angle=-90, width=6.8cm]{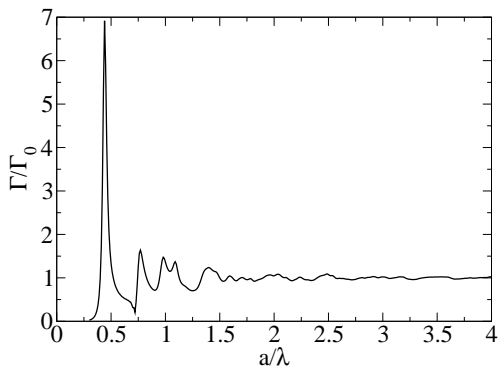} 
\par\end{centering}

\caption{Normalized decay rate of a crystalline structure for a system with
edge of $23$ particles (total system with $528$ particles). The maximum
can be found at $\frac{a}{\lambda}=0.44$.\label{fig:crystalline_spectrum}}
\end{figure}

Results presented in Fig. (\ref{fig:crystalline_spectrum}) for a system with lateral dimension $D=23$ show how it is possible to identify a maximum at $\frac{a}{\lambda}=0.44$ with value $\Gamma/\Gamma_{0}\sim7$. 

\begin{figure}[H]
\begin{centering}
\includegraphics[angle=-90, width=8cm]{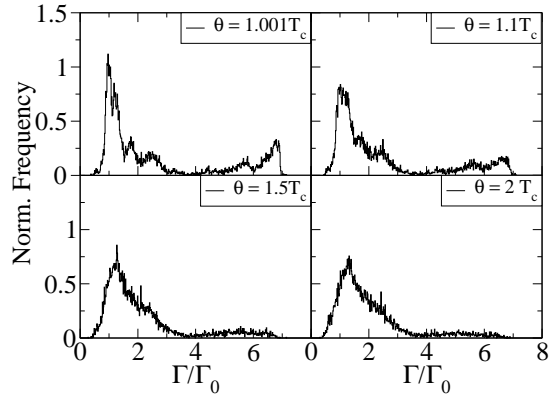} 
\par\end{centering}

\caption{Normalized histogram of the decay rate for $10^{6}$ different configurations
for a system with $D=23$.
\label{fig:temperature}}
\end{figure}

Next, we fix the lattice parameter to $a=0.44\lambda$ and we analyze the decay rate distribution for disordered systems generated at different ordering temperatures ranging from $\theta>>T_{c}$ ($\xi \rightarrow 0$) corresponding to a short-range correlated disorder to $\theta=T_{c}$ ($\xi \rightarrow \infty$), corresponding with a long-range correlated distribution of the vacancies. Results, for $10^{6}$ different configurations, are shown in Fig. (\ref{fig:temperature}).

When the ordering temperature is large ($\theta \sim 2T_{c}$), we obtain a long-tailed distribution centered at $\Gamma/\Gamma_{0} \sim 1.3$, where some rare events are detected for values as high as  $\Gamma/\Gamma_{0} \sim 7$. Similar results have been previously reported \cite{Sapienza_2011, Froufe_2008}.
However, as the ordering temperature decreases towards $T_{c}$, the correlation function between vacancies changes from an exponential decay to a power law  and the decay rate distribution reshapes dramatically. For $\theta=1.001T_{c}$, there is  no longer any long-tailed behavior, but a bimodal distribution where the previously reported rare events increase considerably to build a new maximum centered at $\Gamma/\Gamma_{0} \sim 7$.    

\begin{figure}[H]
\begin{centering}
\includegraphics[angle=-90, width=8cm]{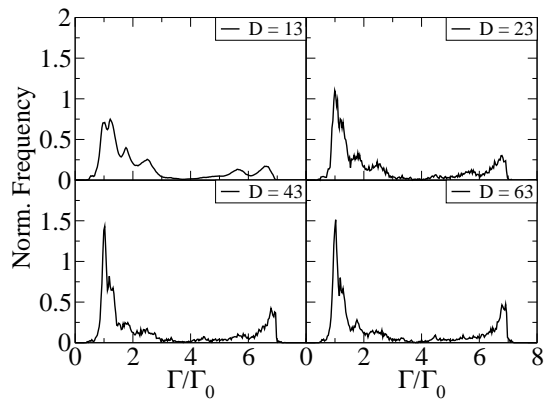} 
\par\end{centering}

\caption{Normalized histogram of the decay rate for $10^{6}$ different configurations.
The systems under study have an edge of $13$, $23$, $43$ and $63$ particles.\label{fig:histogram_all}}
\end{figure}

We have also analyzed the possible presence of finite size effects by considering different lateral sizes ranging from $D=13$ to $D=63$
for $\theta=T_{c}$. Results are shown in Fig. (\ref{fig:histogram_all}). Note how all secondary maxima between $\Gamma/\Gamma_{0} \sim 1$ and $\Gamma/\Gamma_{0} \sim 7$ tend to smear out, while maxima at $\Gamma/\Gamma_{0} \sim 1$ and $\Gamma/\Gamma_{0} \sim 7$ are reinforced when the lateral size increases.

\begin{figure*}
\begin{centering}
\includegraphics[width=0.8\paperwidth]{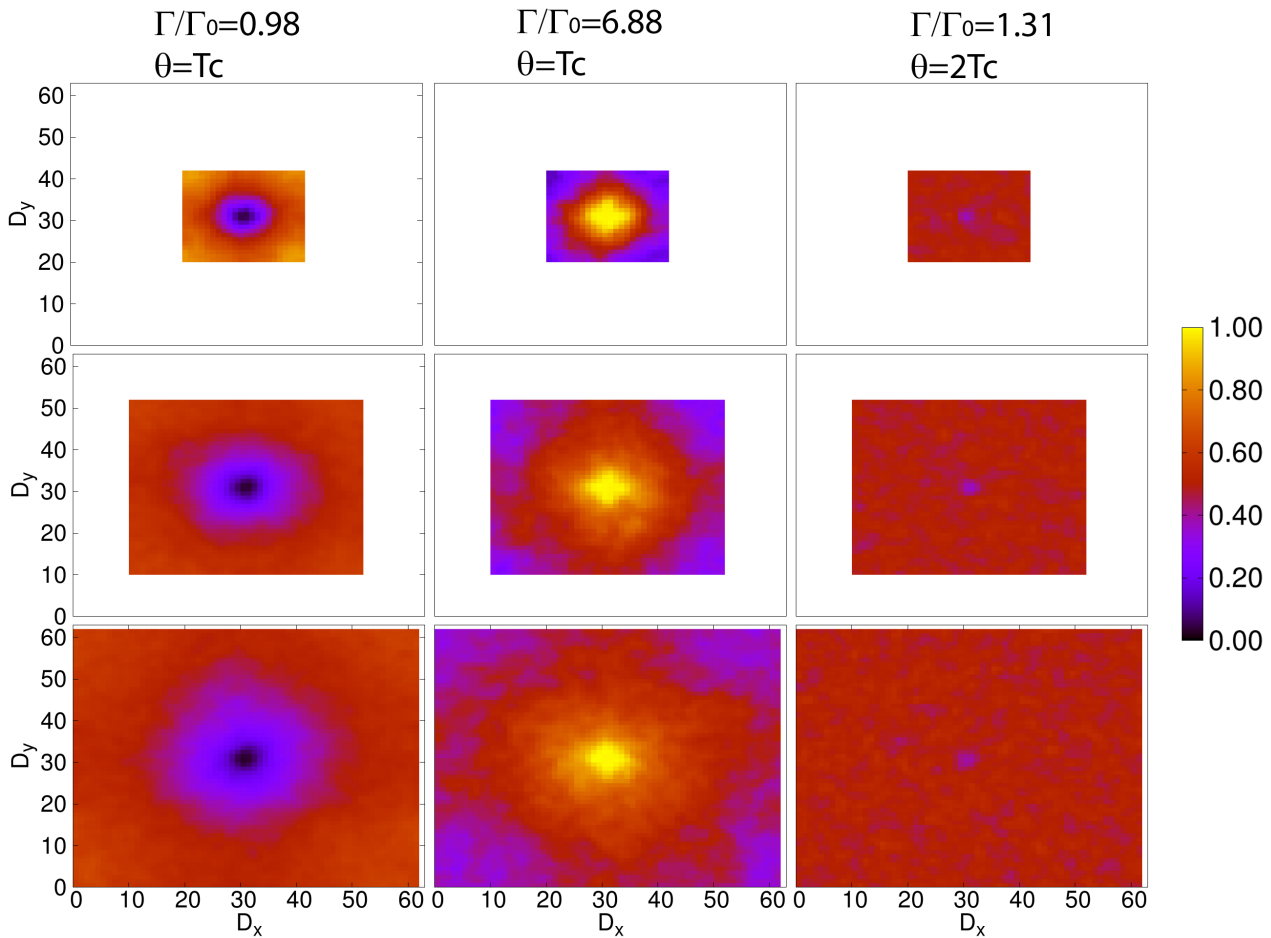} 
\par\end{centering}

\caption{(Color online) Colormap histogram of the occupied positions for different size systems
(on vertical: $D=23$, $D=43$, $D=63$) at different normalized decay
rates and different ordering temperatures. Each histogram is normalized by the number of structures corresponding to each decay rate. \label{fig:col_hist}}
\end{figure*}

Large Purcell factors are due to optical modes confined around the source and are sustained by near-field interactions \cite{Sapienza_2011}, so systems with strong correlations, like the ones reported in this paper, should promote an increase in the number of configurations where the emitting dipole is surrounded by clusters of dipoles allowing for field confinement. To further analyze this idea, we have plotted the histogram of the frequency of occupation for each position normalized to the number of configurations, for different values of the decay rate. We have focused our attention on $\frac{\Gamma}{\Gamma_{0}}=1.31\pm0.02$, corresponding to the maximum of the distribution for $\theta=2T_{c}$ and $\frac{\Gamma}{\Gamma_{0}}=0.98\pm0.02, 6.88\pm0.02$, corresponding to the two main maxima of the distribution for $\theta=T_{c}$ . Results are shown in Fig. (\ref{fig:col_hist}).

For $\frac{\Gamma}{\Gamma_{0}}=1.31$ we detect no special patterns, and dipoles are distributed almost randomly. However, for $\frac{\Gamma}{\Gamma_{0}}=6.88$ (corresponding to the second maximum for $\theta=T_{c}$) the dipoles distribution changes markedly.
In this case, the emitting particle is clearly surrounded by scattering dipoles, allowing for confinement of the optical modes. 
This is in agreement with the attribution proposed in ref. \cite{Sapienza_2011} and clearly shows how large Purcell factors are boosted by long-range correlations in the disordered sample.  Interestingly, for the other maximum located at $\frac{\Gamma}{\Gamma_{0}}=0.98$, the situation is just the opposite and the emitting dipole is, on average, surrounded by vacancies where no field confinement is possible, entailing a dipole response similar to the one found in vacuum. This last effect is also fostered by the existence of long-range correlations when $\theta=T_{c}$. 

In order to understand these effects, it is important to take into account that, at criticality, clusters of all sizes, containing either dipoles or vacancies,  exist on the system and the response due to larger clusters resembles the one found for the crystalline structure $\frac{\Gamma}{\Gamma_{0}} \sim 7$. However, when a non-correlated distribution of vacancies is considered, the correlation length and the cluster's sizes are very small, and the detection of events with a large Purcell factor turns out to be very unlikely.  
The surface structure of the clusters in the Ising model for long-range correlation, i.e. at criticality, is known to be fractal and scale invariant \cite{Coniglio_1989}, like the clusters obtained for high filling factors in semicontinuous metal films experiments \cite{Krachmalnicoff_2010}. These structures are responsible for surface plasmon localization leading to a large increase in the decay rates.

In conclusion, the normalized fluorescent decay rate distribution has been analyzed in a thermally disordered two dimensional diluted dipole lattice where the correlation between vacancies may be tuned at will. When the ordering temperature is far from criticality, the correlation length is small, and the decay rate distribution has the typical long-tailed shape where events with large Purcell factors are extremely rare. However, when the ordering temperature is close to criticality, the correlation length tends to infinity, turning the decay rate function into a bimodal distribution where large Purcell factor events are much more probable.
Our analysis shows how diluted systems, where vacancies are distributed in a long-range manner, enclose clusters of dipoles of all sizes, allowing for the existence of many configurations with optical modes confined around the source. It is worth mentioning the analogy between our model system and  related models in the field of optical lattices filled with cold atoms \cite{Bloch_2008_RMP, Froufe_2009_rlaser, Sherson_2010, Bloch_2012}, in particular for sparsely filled lattices.  This analogy could open another way to study spatial correlations effects on the statistics of fluorescence lifetime.

\begin{acknowledgments}
We acknowledge Dr. Riccardo Sapienza for useful and stimulating discussions. The authors acknowledge the funding from the Spanish Ministry of Economy and Competitiveness through grants "FUNCOAT" CONSOLIDER CSD2008-00023,  "MAPS" MAT2011-29194-C02-01,  "NANOPLAS+"  FIS2012-31070 and "MINIELPHO" FIS2012-36113-C03, and from the Comunidad de Madrid through grant  "MICROSERES-CM" S2009/TIC- 1476 and "NANOBIOMAGNET" S2009/MAT-1726 . L.S.F.-P. acknowledges support from the European Social Fund and CSIC through a JAE-Doc grant. J.J.S. acknowledges  an IKERBASQUE Visiting Fellowship.
\end{acknowledgments}


\end{document}